\documentclass[onecolumn,notitlepage,aps,superscriptaddress]{revtex4-1}
\usepackage{tikz}
\usetikzlibrary{shadings}

\usepackage{cmap}
\usepackage{color}
\usepackage[english]{babel}
\usepackage[utf8]{inputenc}
\usepackage{hyperref}
\usepackage{graphicx}
\usepackage{amsmath}
\usepackage{amssymb}
\usepackage{amsfonts}
\usepackage{amsthm}
\usepackage{textcomp}
\usepackage{scrpage2}
\usepackage[braket]{qcircuit}
\usepackage{verbatim}
\usepackage{latexsym}
\usepackage{pifont}

\newcommand{\bbs}[1]{\boldsymbol{#1}}

\newcommand{\ph}{{\mathrm{sim}}}

\newcommand{\ee}{|\widetilde{E}\rangle}
\newcommand{\eperp}{|\widetilde{E}^\perp\rangle}
\begin{document}
\title{Quantum simulation by qubitization without Toffoli gates}
\author{Mark Steudtner}\affiliation{Instituut-Lorentz, Universiteit Leiden, P.O. Box 9506, 2300 RA Leiden, The Netherlands}
\affiliation{QuTech, Delft University of Technology, Lorentzweg 1, 2628 CJ Delft, The Netherlands}
\author{Stephanie Wehner}
\affiliation{QuTech, Delft University of Technology, Lorentzweg 1, 2628 CJ Delft, The Netherlands}
\begin{abstract}
Qubitization is a modern approach to estimate Hamiltonian eigenvalues without simulating its time evolution.  While in this way approximation errors are avoided,  its resource and gate requirements are  more extensive: qubitization requires additional qubits to store information about the Hamiltonian, and Toffoli gates to probe them throughout the routine.  Recently, it was shown that storing the Hamiltonian in a unary representation can alleviate the need for such gates in one of the qubitization subroutines. Building on that principle, we develop an entirely new decomposition of the entire algorithm: without Toffoli gates, we can encode the Hamiltonian into qubits within logarithmic depth.
\end{abstract}
\date{\today} \maketitle
\maketitle
\section{Introduction}
In digital quantum simulation, a quantum computer acts as a universal simulator for systems difficult to predict with classical methods.  However, the goal of this field is beyond  simple imitation of one system by another: after the Hamiltonian of the  model is   mapped onto qubits, quantum algorithms are employed to extract its spectrum and eigenstates.  The perhaps most sophisticated  algorithm of such kind is quantum phase estimation, which allows one to project into spectral eigenstates by a Fourier analysis of the simulated time evolution (under the model Hamiltonian). However, despite being conceptually simple, quantum phase estimation is challenging on a technical level. Not only are its requirements beyond the abilities of current hardware, but it is likely to even present technical challenges in the future. Part of the problem is that the time evolution cannot be simulated exactly, but must generally be approximated. As originally suggested in \cite{lloyd1996universal}, this can be achieved with Trotterization, which means that the simulator is made to evolve in stroboscopic slices of the exact time evolution.  The shorter the time period of the evolution, the more accurate is the approximation,  but quantum phase estimation has a better resolution for longer time evolutions \cite{suzuki1990fractal,suzuki1991general}.  The algorithm also requires an additional register of estimator qubits to  couple to  every slice in the trotterized time evolution, which is likely to demand nonlocal operations inside the quantum computer. \\

However,  there are more advanced methods that could replace Trotterization in the phase estimation algorithm. In \textit{qubitization} \cite{low2019hamiltonian},  the simulator is extended by a certain number of qubits. The time evolution is then replaced by a unitary that,  in a certain subspace of the extension,  acts as the Hamiltonian on the simulator qubits.  As the unitary describes a rotation out of that subspace, the rotation angles --   functions of the Hamiltonian eigenvalues -- can be read out by the phase estimation routine. The appeal of qubitization is that it does not involve any  approximation of the Hamiltonian; however, it generally  requires higher-level quantum operations, such as the Toffoli gate \cite{poulin2018quantum}. This is in particular the case for when one tries to keep the number of additional qubits to a minimum. For qubitization, Toffoli gates can be regarded as a symptom of a compressed data structure, meaning that bit strings are encoded nonlinearly. For the unitary entangling the simulator qubits to the extension, Toffoli gates signify a drastic increase in circuit depth. Unfortunately,  the depth of a quantum algorithm (its theoretical runtime)  is a bottleneck of every quantum computation. \\

 Recently, a qubitization version with a decompressed storage structure was introduced in \cite{poulin2018quantum}, opening an important subroutine of the algorithm up for parallelization. Here, we want to build upon this work, and obtain a qubitization version that is entirely free of Toffoli gates. For that purpose, we apply  recently popular   state preparation techniques \cite{izmaylov2019unitary,bonet2019nearly},  and so parallelize the remaining subroutines of  qubitization. The new routines also fit  more naturally into the framework of quantum  phase estimation:  rather than being in permanent contact  with the estimator qubits like the Trotter circuits, the qubitization routines have relay points, at which singular controlled gates  can act as switches. Without any compression, we will present a version of qubitization in which all its components are of  low  algorithmic depth. For local Hamiltonians,  this phase estimation version  has a total runtime  scaling logarithmically with the number of terms.

\section{Background}
\label{sec:background}
The goal of quantum simulation is to extract the eigenstates and eigenvalues of a Hamiltonian acting on qubits. Let that Hamiltonian have $\Lambda$ terms and be  of the form
\begin{align} \label{eq:hamiltonian}
H = \sum_{k=1}^\Lambda \alpha_k \cdot  p^k_\ph \, ,
\end{align}
where $ p^k_\ph$ are Pauli strings --  signed products of Pauli operators $\lbrace X, \, Y, \, Z \rbrace$ acting on different qubits  inside the simulator, a quantum register we will call `$\ph$'.  By shifting all the minus signs into the Pauli strings,   their coefficients $\alpha_k$  are positive and  for the moment we will normalize them to $\sum_{j=1}^{\Lambda} \alpha_j \leq 1$, such that $-1\leq  E \leq  +1$ for all energy eigenvalues $E$.   Trotterized time evolution and other product formulas \cite{suzuki1990fractal,suzuki1991general,poulin2014trotter,wecker2014gate,childs2019faster,kivlichan2019phase} approximate $\mathrm{exp}(iH)$ by products of Pauli string rotations $\exp(i\alpha_k\, p_\ph^k \, \Delta )$  for different time slices $\Delta\leq1$. Rotations of Pauli strings are generally straightforward to implement \cite{nielsen2000quantum}.   Using those textbook circuits, quantum phase estimation \cite{cleve1998quantum} can  discern  states with different eigenvalues $E$.  The projective power of the phase estimation algorithm is generally bigger the more qubits it has at its disposal.
In its  minimal version \cite{kitaev1995quantum},  Kitaev's phase estimation  (shown in  Figure \ref{fig:circs}(a)), only one estimator qubit is required:  Hamiltonian eigenvalues are inferred from its measurement histogram.   The routine in  Figure \ref{fig:circs}(a)  features the component $U$, an approximation to the simulated time evolution, $U \approx \exp(iH)$. \\

An alternative to trotterized time evolution is qubitization \citep{low2019hamiltonian}. In qubitization without quantum signal processing \cite{low2017optimal}, not the energy $E$ is estimated, but $\pm \arccos(E)$, where the  sign $\pm$  occurs for every run of the circuit at random.  While Trotterization only approximates the time evolution $e^{iH}$, qubitization employs a sequence of  two unitaries, $\widehat{V}$ and $\widehat{S}$, to estimate   the phase factors $\exp(i \arccos E) = E\pm i\sqrt{1-E^2}$ for the exact energy eigenvalues $E$.

However, qubitization has a major drawback:  it  requires an additional register of at least $n \geq \log \Lambda$ qubits. We will refer to it as the qubitization register. In those qubits,  information about the Hamiltonian must be encoded with a unitary $\widehat{G}$. This operation prepares a state  $\ket{G}$, which features the weights of the Hamiltonian $\lbrace\sqrt{\alpha_k}\rbrace$. Besides $\widehat{S}$, a reflection of  $\widehat{G}$,  $\widehat{S}\ket{G} = -\ket{G}$, an entangling operation $\widehat{V}$ is required to facilitate the action of the Hamiltonian (on the simulation register)  within the subspace of $\ket{G}$, i.e.~$\bra{G}\widehat{V}\ket{G}=H$.  All three unitaries have the form \cite{low2019hamiltonian}:
\begin{align}\label{eq:old_unitaries1} \widehat{G} \ket{\bbs{0}} &= \ket{G} = \sum_{k=1}^{\Lambda} \sqrt{\alpha_k} \ket{\bbs{\mu_k}}\, , \\
 \label{eq:old_unitaries2} \widehat{S} &= 2\ket{G}\!\!\bra{G}- \mathbb{I}\, , \\
 \label{eq:old_unitaries3}
\widehat{V} &= \sum_{k=1}^{\Lambda}  \ket{\bbs{\mu_k}}\!\!\bra{\bbs{\mu_k}} \otimes p_\ph^k \; +\;  \ket{\bbs{0}}\!\!\bra{\bbs{0}}\otimes \mathbb{I}_\ph \; + \;  \text{\dots} \; ,
\end{align}
where $\ket{\bbs{0}}$ is the state in which all qubits are in $\ket{0}$, and   $\lbrace\ket{\bbs{\mu_k}}\rbrace$  is a subset of the computational basis not containing $\ket{\bbs{0}}$.  The state $\ket{G}$ in \eqref{eq:old_unitaries1} superposes the configurations $\ket{\bbs{\mu_k}}$ according to the weights $\lbrace \alpha_k \rbrace$,  which from now on shall be normalized to
$\sum_k \alpha_k=1$. In \eqref{eq:old_unitaries3}, only the relevant part of  $\widehat{V}$ is presented: its action on the $\ph$-register given it finds the states   $\lbrace\ket{\bbs{\mu_k}}\rbrace$ in the qubitization register, and trivial action if the qubitization register is in $\ket{\bbs{0}}$.  The action of  $\widehat{V}$ outside of this subspace does not play a role in the qubitization process. Note that throughout the literature, the unitaries $\widehat{G}$ and $\widehat{V}$ are also referred to as $\textsc{prepare}$ and $\textsc{select}$ \cite{babbush2018encoding}.\\
A na\" ive implementation of the qubitization routine would now replace the controlled time evolution, $U$, in Figure \ref{fig:circs}(a) with the sequence of controlled $\widehat{V}$ and $\widehat{S}$. However,  $\widehat{V}$  features controlled gates already, and so the amount of Toffoli gates would increase if $\widehat{V}$ was applied conditionally. Fortunately, as was noticed in \cite{poulin2018quantum}, $\widehat{V}$ acts trivially if $\ket{G}$ is not initialized, and the same applies to $\widehat{S}$. For the benefit of the gate complexity,  the application of $\widehat{G}$ was made conditional, and $\widehat{G}$ was reintroduced after $\widehat{S}\cdot\widehat{V}$ for only the $\ket{0}$ subspace of the estimator qubit, see Figure \ref{fig:circs}(b). Note that using qubitization, the phase estimation does no longer project into eigenstates of the Hamiltonian (in the $\ph$-register).   When estimating $\pm\arccos E$, the eigenstate of $E$ must be retrieved probabilistically.  Again we learn from \cite{poulin2018quantum}, that the unitary $\widehat{G}$ can be applied once more after the circuit in Figure \ref{fig:circs}(b). A following
 measurement  of  every qubit  in the  qubitization register  has then  a  $50 \%$ chance of all outcomes  being $+1$, which is a flag that the $\ph$-register has been projected into the eigenstate of $E$.

Within this work, we are particularly interested in the basis of the qubitization register. Here,  \\ $\bbs{\mu_k}= \mu_{k,1} \; \mu_{k,2}\;  \mu_{k,3}\; \cdots \; \mu_{k,n}$
 is a string of $n$ bits, $\mu_{k,j} \in \lbrace 0,1\rbrace$, describing a computational basis state by $\ket{\bbs{\mu_k}} = \bigotimes_{j=1}^n \ket{\mu_{k,j}}_j.$ The set of bit strings  $\lbrace \bbs{\mu_k}\rbrace$  is determined by the encoding (i.e.~the mapping: $k\mapsto \bbs{\mu_k}$) which  has important implications for the depth of $\widehat{G}$, $\widehat{S}$ and $\widehat{V}$. Prior works have considered two encoding strategies. \\

\textit{Binary encoding} \cite{low2019hamiltonian,babbush2018encoding}  --  To reach the minimal qubit requirements of $n=\log\Lambda$, all possible bit strings  $\lbrace 0,1 \rbrace ^{\otimes n}$ must be considered as basis configurations $\lbrace \bbs{\mu_k} \rbrace$. For the construction of  $\widehat{V}$,  sequences of  multi-qubit Toffoli gates must probe  the qubitization register for  configurations $\bbs{\mu_k}$ and  apply   $p^k_\ph$  for all $k$ from $1$ to $\Lambda$, see Figure \ref{fig:reps}(a) or \cite{babbush2018encoding}. A comparable effort is required when constructing the state $\ket{G}$, where   $ O(\Lambda)$ rotations have to be conditioned on specific basis configurations, probed again with Toffoli gates.  The algorithmic depth of $\widehat{V}$, $\widehat{S}$ and $\widehat{G}$ is therefore fixed to be at least  proportional to $\Lambda$.\\

\begin{figure}
{ \centering
\begin{tikzpicture}
   \node[] at (0,0) {\Qcircuit @C=1.4em @R=.2em {\lstick{} & 	 \multigate{4}{\widehat{G}^\dagger}& 	\ctrlo{1}	& \multigate{4}{\widehat{G}}	\qw & \qw	\\
\lstick{} & 	\ghost{\widehat{G}^\dagger} &	\ctrlo{1}&	\ghost{\widehat{G}} &	\qw &	\\
\lstick{} & 	\ghost{\widehat{G}^\dagger} &	\ctrlo{1}&	\ghost{\widehat{G}} &	\qw &	\\
\lstick{} & 	\ghost{\widehat{G}^\dagger} &	\ctrlo{1}&	\ghost{\widehat{G}} &	\qw &	\\
\lstick{} & 	\ghost{\widehat{G}^\dagger} &	 \controlo  \qw &	\ghost{\widehat{G}} &	\qw &	} };
\end{tikzpicture}
\caption{Implementation of $\widehat{S}$ with a Toffoli-type gate  in a 5-qubit qubitization register \cite{poulin2018quantum,babbush2018encoding}. When using a unary encoding, the size of this register will be proportional to the number of Hamiltonian terms, resulting in a large gate.  }\label{fig:stoffoli}
 \par }
\end{figure}
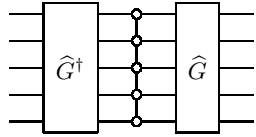

\textit{Unary encoding} \cite{poulin2018quantum}  -- To eliminate the need for Toffoli gates in the construction of $\widehat{V}$ (but not directly $\widehat{S}$), we can encode information in $\ket{G}$ with only the strings $\bbs{\mu_k} = \bbs{e_k}$, where $\bbs{e_k}$ is the bit string corresponding to the $k$-th unit vector with $e_{k,j}=\delta_{jk}$. While this increases the qubit count to $n=\Lambda$, it allows one to define   $\widehat{V}$ in  a simpler way
\begin{align}\label{eq:shallowV}
\widehat{V} = \prod_{k=1}^{\Lambda} \left(\ket{0}\!\!\bra{0}_k + \ket{1}\!\!\bra{1}_k  \otimes p^k_\ph \right)\, ,
\end{align}
which means that instead of checking the states of the entire qubitization register, each string $p^k_\ph$ is applied conditioned on the $k$-th qubit for all $k$ between $1$ and $\Lambda$, see Figure \ref{fig:reps}(b).  Since each of those gates is controlled by a different qubit, the only  thing hindering a massive parallelization of $\widehat{V}$ is the structure of the $p$ strings themselves:  the complexity of \eqref{eq:shallowV} depends on their individual Pauli weight and whether they overlap with one another. Fortunately, there is a considerable body of theoretical work on  how to produce local strings when simulating fermions  \cite{bravyi2002fermionic,seeley2012bravyi,havlivcek2017operator,setia2018superfast,steudtner2019quantum,jiang2018majorana}  and bosons \cite{sawaya2019resource}.   While the decompressed data structure of the unary encoding benefits the  complexity of $\widehat{V}$,  benefits for the other components have not been shown: regardless of the encoding,   $\widehat{S}$  is suggested to be implemented by the sequence   $\widehat{G}\cdot(2 \ket{\bbs{0}}\!\!\bra{\bbs{0}}-\mathbb{I})\cdot\widehat{G}^\dagger$   \cite{babbush2018encoding,poulin2018quantum}, with the reflection $(2 \ket{\bbs{0}}\!\!\bra{\bbs{0}}-\mathbb{I})$ requiring a Toffoli-type gate across the entire qubitization register, see Figure \ref{fig:stoffoli}. For the unary encoding, this gate would have particularly many controls -- the $\Lambda$-fold Toffoli would most certainly dominate the time complexity of the entire algorithm. Here, adding more qubits would help:  a depth of $\log \Lambda$ could theoretically be achieved by doubling the size of the  qubitization register, however the high resource requirements are a critical downside of the unary encoding already.  $\Lambda$ must typically  be regarded as a large number, such that adding another $\Lambda$ qubits is a serious repercussion. Other strategies with which the reflection could be approximated can be found in \cite{chowdhury2018improved}.  Fortunately, the unary encoding does not actually require  $\widehat{S}$  a high-level gate and the following sections will replace $\widehat{S}$ and $\widehat{G}$ with a series of Pauli string rotations.

\begin{figure}
{ \centering
\begin{tikzpicture}
\node[] at (-3,1.5){\textbf{(a)}};
\node[] at (4,1.5){\textbf{(b)}};
\node[] at (8,0) {
\Qcircuit @C=1.4em @R=1.2em {
 \lstick{\ket{+}_{\text{est}}} & \qw &	\ctrl{1} &\qw &	\qw & 	\ctrlo{1} & 	\gate{\mathrm{H}} & 	\meter	\\
 \lstick{\ket{\bbs{0}}} &\qw / &	\gate{\widehat{G}} &	\multigate{1}{\widehat{V}} &	\gate{\widehat{S}}&  \gate{\widehat{G}} &	\qw 		\\
\lstick{\ket{\varphi}_\ph} & \qw /&	\qw &	\ghost{\widehat{V}} &	\qw	& \qw & \qw }
};

\node[] at (0,0){\Qcircuit @C=1.4em @R=1.2em {
 \lstick{\ket{+}_{\text{est}}} &		\ctrl{1} & 	\gate{\mathrm{H}} & 	\meter 	\\
\lstick{\ket{\varphi}_\ph} &		\gate{U} &	\qw  }
};

\end{tikzpicture}
\par }
\caption{Quantum phase estimation circuits. $\mathrm{H}$ is the Hadamard gate and $\ket{\varphi}_\ph$ a trial state of the simulator. \textbf{(a)} Kitaev's phase estimation circuit \cite{kitaev1995quantum}. $U$ is an approximation to the simulated time evolution $\mathrm{exp}(iH)$. \textbf{(b)} Qubitization circuit \cite{poulin2018quantum} with one estimator qubit, featuring the three unitaries $\widehat{V}$,  $\widehat{S}$ and  $\widehat{G}$, where only the latter is applied conditionally. All qubitization qubits are initialized in the zero state, such that the entire register equals $\ket{\bbs{0}}=\bigotimes_{j=1}^n\ket{0}_j$.}\label{fig:circs}
\end{figure}
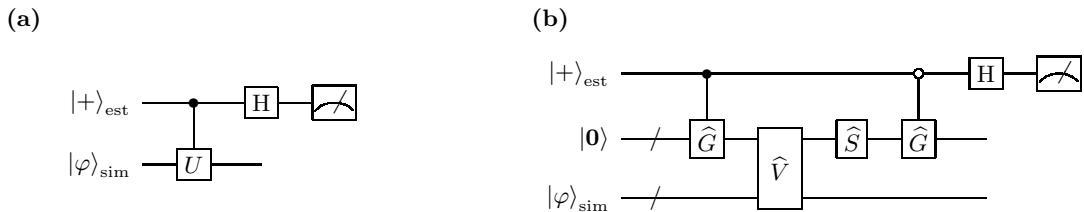
  \begin{figure}
   { \centering

  \begin{tikzpicture}
  \node[] at (0,0){
  \Qcircuit @C=1.4em @R=1.2em {
\lstick{1}&\qw&	\ctrl{1}&	\ctrl{1}&	\ctrlo{1}&	\ctrlo{1}&	\ctrl{1}&	\qw &	\\
\lstick{2}&\qw&	\ctrlo{1}&	\ctrl{1}&	\ctrl{1}&	\ctrl{1}&	\ctrl{1}&	\qw &	\\
\lstick{3}&\qw&	\ctrlo{1}&	\ctrlo{1}&	\ctrlo{1}&	\ctrl{1}&	\ctrl{1}&	\qw &	\\
\lstick{\ph}&\qw /&	\gate{p^1}&	\gate{p^2}&	\gate{p^3}&	\gate{p^4}&	\gate{p^5}&	 \qw 	\\
  }};

    \node[] at (7.5,0){
  \Qcircuit @C=1.4em @R=1.2em {
  \lstick{1}&\qw&	\ctrl{4}	\qw &	\qw &	\qw &	\qw & \qw 	\\
\lstick{2}&\qw&	\qw &	\ctrl{3}	\qw &	\qw &	\qw & \qw  	\\
\lstick{3}&\qw&	\qw &	\qw &	\ctrl{2}	\qw &	\qw & \qw 	\\
\lstick{4}&\qw&	\qw &	\qw &	\qw &	\ctrl{1}	\qw &  \qw 	\\
\lstick{\ph}&\qw /&	\gate{p^1}&	\gate{p^2}&	\gate{p^3}&	\gate{p^4}&	\qw & 	\\
  }};

     \node[] at (-4,1.4){\textbf{(a)}};
      \node[] at (4,1.4){\textbf{(b)}};
  \end{tikzpicture}
  \par }
  \caption{ Implementations of $\widehat{V}$. \textbf{(a)} Binary representation requiring $n=\log \Lambda$ qubits. Sequences of multi-Toffoli gates probing the state $\ket{G}$ for the conditioned application of singular $p^k$. To benefit from cancellations between adjacent gates \cite{babbush2018encoding}, $\bbs{\mu_k}$ is the $k$-th word of the Gray code \cite{sawaya2019resource}. However, even with those cancellations, the depth of $\widehat{V}$ is $O(\Lambda)$.   \textbf{(b)} Unary encoding requiring $n=\Lambda$ qubits. The application of the string $p^k$ is only controlled on the $k$-th qubit in the qubitization register. Since non-overlapping $p$ strings can be applied in parallel, $\widehat{V}$ can in the best case be implemented within $O(1)$ time. } \label{fig:reps}
  \end{figure}
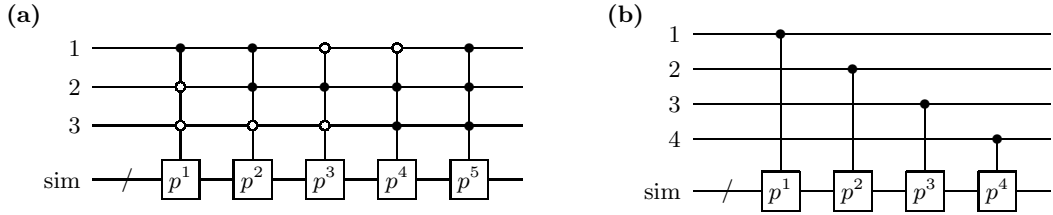

\section{Results}

\label{sec:results}

In this section, we  summarize the techniques and findings  of this work, and give an overview of its further organization.
This work studies  generalized unary encodings in quantum simulation with qubitization: for these encodings, the unitaries \eqref{eq:old_unitaries1} - \eqref{eq:old_unitaries3} are defined with a set of configurations $\lbrace \bbs{\mu_k} \rbrace$ that are linearly independent. We here eliminate Toffoli gates from the qubitization circuits of  these  encodings. This means in particular that we simplify $\widehat{S}$, for which a Toffoli-type gate like in Figure \ref{fig:stoffoli} is not necessary.  Instead, we decompose the routines $\widehat{G}$ and $\widehat{S}$ into rotations of Pauli strings, such that the implementation of a qubitization routine becomes as straightforward as trotterizing a time evolution. Pauli string rotations  can be further decomposed into a sequence of two-qubit gates and single-qubit rotations with an up to  logarithmic scaling in the number of qubits  involved \cite{motzoi2017linear}.     \\

We will now give a high level overview of our results.  The decomposition is  achieved in two stages: first  $\widehat{G}$ and $\widehat{S}$ are expressed as linear combinations of Pauli strings -- the correctness of these decompositions is proven in Section \ref{sec:proof}. Then, we review gadgets with which these linear combinations can be created. This is done in Section \ref{sec:techniques}.  The simplification of the qubitization circuit is however not the only benefit of the decomposition:  in Section \ref{sec:logdepth} we introduce a particular encoding with $\bbs{\mu_k}\neq \bbs{e_k}$,  for which we can implement $\widehat{G}$ and $\widehat{S}$ within $O(\log \Lambda)$ depth. Discussions about the qubit connectivity are shifted to the conclusion  of this work (Section \ref{sec:conclusion}).   For the remainder of this section, we will describe our findings analytically, and integrate them into the phase estimation routine before giving an overview of the entire protocol.

Before we start, however,  we would like to introduce a few shorthands  used  throughout this paper: let  us write the set of integers from $1$ to $N$ as $[N]$ and $[0]=\emptyset$.  We define sums over an empty set to yield zero and products over an empty set to yield one. Also, we will index Pauli operators with the qubits they act on, e{.}g{.} $Z_1$, $X_2 \otimes Y_4$.  \\

Rather than Toffoli gates, $\widehat{G}$ and $\widehat{S}$ are going to be built from a set of  Pauli strings $\gamma_j^x$, $\gamma^y_j$, for all $j \in [\Lambda] $. All of these $2\Lambda$ Pauli strings anticommute pairwise -- they are in fact qubit representations of  Majorana fermions. For $\bbs{\mu_k} = \bbs{e_k}$, the strings are of the form
\begin{align}\label{eq:majorana}
\gamma_j^x \; = \;  \left(\bigotimes_{k\in [j-1]} Z_k \right) \otimes X_j\qquad \text{and} \qquad
\gamma_j^y \; = \;\left(\bigotimes_{k\in [j-1]} Z_k \right)  \otimes Y_j \, ,
\end{align}
which would correspond to Jordan-Wigner transforms of their fermionic selves. Note however that these strings are only tools, which,  living in the qubitization register, have no  physical meaning in the simulated model. The strings are chosen such that they can create the basis states by $\gamma^x_k \ket{\bbs{0}} = -i \gamma^y_k \ket{\bbs{0}} = \ket{\bbs{\mu_k}}$, and their anticommutation properties are desired such that the linear combinations
\begin{align}\label{eq:gee1}
g^x \; = \; &  \sum_{j \in [\Lambda]} \sqrt{\alpha_j} \gamma_j^x \, \quad \text{and} \\ \label{eq:gee2}
g^y\; = \; \; &  \sum_{j \in [\Lambda]}  \sqrt{\alpha_j} \gamma_j^y \, ,
\end{align}
are unitary. The operators \eqref{eq:gee1} and \eqref{eq:gee2} can then be applied  directly to the system through gadgets based on Pauli string rotations.  This means we can construct the operators $\widehat{G}$ and $\widehat{S}$ as we find them to be made up by $g^x$ and $g^y$:
\begin{align} \label{eq:results}
\widehat{G} = g^x\, ,  \qquad \widehat{S} = -i g^y \cdot g^x\, .
\end{align}
 While $\widehat{G}$ can be prepared by a single instance of the aforementioned gadgets, the product within $\widehat{S}$ means that it is implemented by two consecutive gadgets. Not only can we build the linear combinations $g^x$ and $g^y$, but also their rotations and conditional versions, e{.}g{.} $\mathrm{exp}(i\theta g^x) = \cos\theta+i\sin\theta\cdot g^x$ and $\ket{0}\!\!\bra{0}_{\mathrm{est}}\otimes \mathbb{I}+\ket{1}\!\!\bra{1}_{\mathrm{est}}\otimes g^y $, such that the gadgets can be integrated easily into the phase estimation routine.   \\
Using the fine structure of $\widehat{S}$ and assuming we additionally apply the gadget for  $g^x$ (just like in the trick of \cite{poulin2018quantum}, reclaiming the Hamiltonian eigenstates), we present the qubitization circuit featuring only gadgets for $g^x$  and $g^y$ in Figure \ref{fig:zerg}. One can easily verify that up to a phase shift, this circuit acts as $\widehat{G} \widehat{S} \widehat{V} \widehat{G}$ in the $\ket{1}$ subspace of the estimator qubit and as the identity in its $\ket{0}$ subspace.   \\
\subsection{Protocol}
The entire process of simulating  a Hamiltonian \eqref{eq:hamiltonian} can be summarized with the following instructions.

\begin{figure}
{ \centering
\begin{tikzpicture}
   \node[] at (0,0) {\Qcircuit @C=1.4em @R=1.2em { \lstick{\ket{+}_{\text{est}}} & \qw	&\ctrl{1} &	\qw &	\ctrl{1} &	\gate{\mathrm{H}} &	\meter	\\
 \lstick{\ket{\bbs{0}}} & \qw /	& \gate{g^x} &	\multigate{1}{\widehat{V}} &	\gate{g^y} &	\qw&	\qw	\\
\lstick{\ket{\varphi}_\ph} &	\qw / &\qw	&\ghost{\widehat{V}} &	\qw &	\qw &	\qw	} };
\end{tikzpicture}
\caption{Quantum phase estimation circuit using qubitization with a unary encoding. This circuit, estimating  $\pm\arccos(H)$ up to a constant phase shift,  features the subroutines $g^y$ and $g^x$ that can be parallelized to the runtime of $O(\log \Lambda)$.  }\label{fig:zerg}
 \par }
\end{figure}
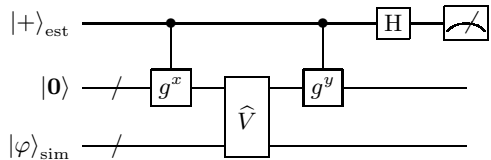

\begin{enumerate}
\item Extract the coefficients $\lbrace\alpha_k\rbrace$ and Pauli strings $\lbrace p^k_\ph\rbrace$ from the Hamiltonian.
\item Choose an encoding:  $k\mapsto \bbs{\mu_k}$ and compute the sets of Pauli strings $\lbrace\gamma_k^x\rbrace$ and $\lbrace\gamma_k^y\rbrace$.
\item Decide the type of gadgets and obtain their parametric angles from the Hamiltonian coefficients:   $\lbrace \alpha_k \rbrace \mapsto \lbrace \phi_k \rbrace$.
\item  Compile the quantum circuits for the gadgets of $g^y$ and $g^x$ according to  $\lbrace \gamma^x_k \rbrace$, $\lbrace \gamma^y_k \rbrace$,  and  $\lbrace \phi_k \rbrace$ as well as the  entangling routine $\widehat{V}$ with respect to the basis  $\lbrace\bbs{\mu_k}\rbrace$  and the set of Pauli strings  $\lbrace p^k_\ph\rbrace$.
 \item Run the circuit in Figure \ref{fig:zerg} on a quantum computer to estimate energy eigenvalues. Measure the qubitization register for a 50\% chance to recover Hamiltonian eigenstates.
\end{enumerate}

\section{Techniques}
\label{sec:techniques}
In this section we will present gadgets for the operators  \eqref{eq:gee1} and \eqref{eq:gee2}  that are essential for a Toffoli-free decomposition of the operators $\widehat{G}$ and $\widehat{S}$. Rather than Toffoli gates, the decomposition relies on sequences of Pauli string rotations, rendering the hardware requirements basically identical to those of trotterized time evolution, for which Pauli string rotations are also the elementary building blocks.  In the most general use case, these gadgets receive a set of $L$ pairwise-anticommuting Pauli strings $\lbrace h^k \rbrace$, as well as a set of $L$ real coefficients $\lbrace \beta_k  \rbrace$  and realize (controlled versions of) the operators
\begin{align} \label{eq:slater_rotation}
\mathrm{exp}\left(i\sum_{k=1}^{L} \beta_k \cdot h^k\right) = \cos \theta + i \sin \theta\cdot \left(\frac{1}{\theta} \sum_{k=1}^{L} \beta_k \cdot h^k \right) \,
\end{align}
where  $\theta = (\sum_k |\beta_k|^2)^{1/2} $.  The  gadgets   contain   $2L-1$ Pauli string rotations featuring some angles $\phi_1, \, \phi_2,  \, \dots\, , \,  \phi_L$, which have to be matched with the coefficients $\lbrace \beta_k \rbrace$ in classical pre-processing.
 Here we consider two different gadgets that we will refer to as  `symmetric' and `antisymmetric', found in \cite{izmaylov2019unitary}  and \cite{bonet2019nearly}, yielding circuits of depth as short as  $O(L)$ and $O(\log L)$, respectively. In both gadgets we identify  relay points  to control their application on an external qubit without adding a control to every gate. For the implementation of $g^x$ or $g^y$, we find $L=\Lambda$, and   $\lbrace h^j \rbrace$ equal to $\lbrace \gamma_j^x \rbrace$ or  $\lbrace \gamma_j^y \rbrace$. However, while we equate these sets, we do not necessarily want to equate all individual elements with the same index, so  $\lbrace h^j \rbrace = \lbrace \gamma_j^x \rbrace$  does not necessarily imply that for all $j$ we have $h^j=\gamma_j^x$.
  Our purposes require $\theta=\pi/2$, so the set of parameters $\lbrace \beta_k \rbrace$ must be equal to $\left\lbrace \sqrt{\frac{\pi}{2}\alpha_k }\right\rbrace$.
\subsection{Symmetric gadget}
\begin{figure}
{ \centering
\begin{tikzpicture}
\node at (-6.6,-4.3) {\textbf{(a)}};
\node at (-6.6,-6.8) {\textbf{(b)}};
    \node[] at (0,-5) {
\Qcircuit @C=.7em @R=1.2em {  \lstick{\text{control}}&\qw  & \qw &  \qw &    \ctrlo{1}  &	\qw  &\ctrlo{1} & \qw   \\ \lstick{\text{target}} & \qw /
& \gate{\text{\vphantom{Rg}Left arm}}
&  \gate{ih^{1}(\phi_{1})}  &  \gate{ih^{1}(\frac{\pi}{2} - \phi_{1})}  & \gate{\text{Right arm}}  & \gate{ih^1(-\frac{\pi}{2})} & \qw }
    };
  \node[right] at (-6.2,-8) {   \Qcircuit @C=1.4em @R=1.2em {
& \gate{\text{Right arm}}  & \qw & \push{ = \quad  \; \; }  & \gate{i h^2(\frac{1}{2}\phi_2)}  &  \gate{i h^3(\frac{1}{2}\phi_3)}& \gate{ih^4(\frac{1}{2}\phi_4)}  & \qw & \cdots &  &\gate{ih^L(\frac{1}{2}\phi_L)}
}};
  \node[right] at (-6.,-7) {   \Qcircuit @C=1.4em @R=1.2em {
& \gate{\text{\vphantom{Rg}Left arm}}  & \qw & \push{ = \quad  \; \; }  & \gate{i h^L(\frac{1}{2}\phi_L)}  &  \qw & \cdots &  & \gate{i h^4(\frac{1}{2}\phi_4)}& \gate{ih^3( \frac{1}{2}\phi_3)}  & \gate{ih^2(\frac{1}{2}\phi_2)}
}};
    \end{tikzpicture}  \par}
    \caption{Symmetric gadget for the linear combination of pairwise anticommuting strings $\lbrace h^k \rbrace$, controlled on a single qubit. The gates labeled $ih^k(\phi)$ denote rotations $\exp (i h^k \phi)$ around angles $\phi$. \textbf{(a)} The main circuit featuring a   $h^1$ rotation, in between the `right arm' and `left arm' subroutines. The control relay is built around  the fact that $h^1\, (\text{Right arm})\, h^1 = (\text{Left arm})^\dagger$. \textbf{(b)} Right and left arm, which are sequences of rotations of the strings $\lbrace h^k \rbrace_{k=2}^L$. Comparing the two arms, the order of their rotations is reversed.} \label{fig:trotter}
\end{figure}
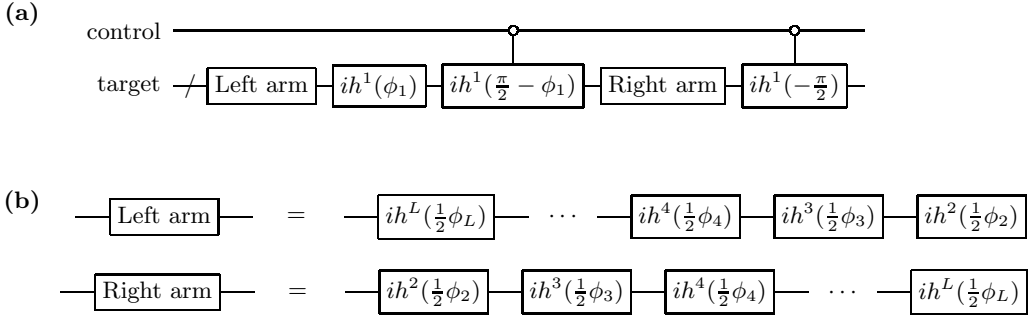

The first version of this gadget works in a very intuitive way. The  symmetric gadget  is based on second order Trotterization of the $h$ strings. Since all those strings anticommute, this Trotter approximation does not yield \eqref{eq:slater_rotation} when setting $\phi_k = \beta_k$ for all $k \in [L]$. However, the result of the Trotterization is still quite predictable, in fact arbitrary   $\lbrace \phi_k \rbrace$ turn out to be the generalized Euler angles in the resulting superposition \cite{hoffman1972generalization}, i{.}e{.}~when  $\lbrace \sqrt{\frac{\pi}{2}}\beta_k\rbrace$ describe the Cartesian coordinates on an $L$-dimensional unit sphere, its spherical coordinates are  $\lbrace\phi_k\rbrace$:
\begin{align} \notag
 & 	\mathrm{exp}\left( \frac{i}{2} \phi_L\cdot h^L\right)\mathrm{exp}\left( \frac{i}{2} \phi_{L-1}\cdot h^{L-1}\right) \cdots \, \mathrm{exp}\left( \frac{i}{2} \phi_{2}\cdot h^{2}\right)  \mathrm{exp}\left( i \phi_{1}\cdot h^{1}\right)  \\ & \times \; \mathrm{exp}\left( \frac{i}{2} \phi_{2}\cdot h^{2}\right) \cdots \, \mathrm{exp}\left( \frac{i}{2} \phi_{L-1}\cdot h^{L-1}\right)  \mathrm{exp}\left( \frac{i}{2} \phi_L\cdot h^L\right)  \notag \\
 & =  \; \prod_{m \in [L]} \cos \phi_m \; +\;  i \sum_{k\in [L]} h^k \cdot \sin\phi_k \prod_{j \in [k-1]} \cos \phi_j \, .
  \end{align}
  One can now arrive at \eqref{eq:slater_rotation} by setting $\phi_1 = \arcsin(\frac{\beta_1}{\theta} \sin \theta)$  and then progress iteratively with\\  $\phi_{m+1} = \arcsin \left( \frac{\beta_{m+1}}{\beta_m}  \tan(\phi_m)  \right)$.  As shown in Figure \ref{fig:trotter}, the gadget can be switched off from an external control qubit  by completing the innermost $h^1$ rotation to $\pi/2$.  Canceling the right and left arm of the Trotterization,  the circuit becomes trivial with one more controlled application of $(-ih^1)$ on the outside.

  \subsection{Antisymmetric gadget}
  \label{subsec:anti}
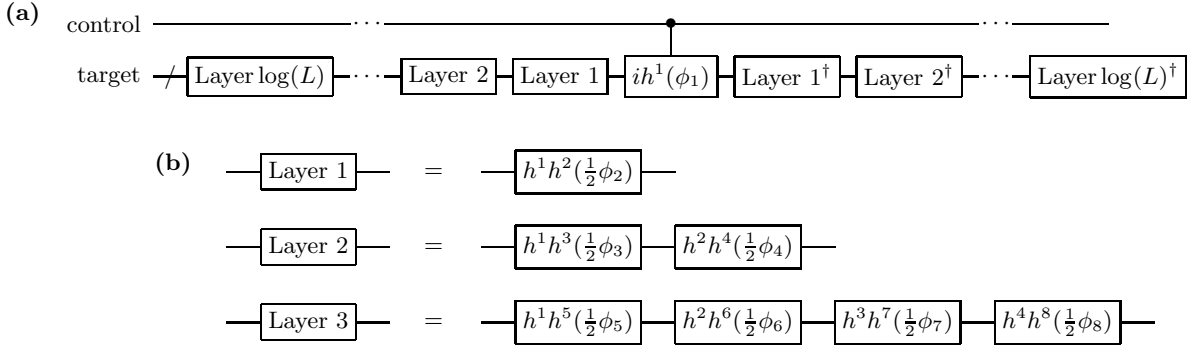
\begin{figure}
{ \centering

\begin{tikzpicture}
\node[] at (-8.6,.6) {\textbf{(a)}};
\node[] at (-6.6,-1.4) {\textbf{(b)}};
\node[] at (0,0) {  \Qcircuit @C=.7em @R=1.2em {
  \lstick{\text{control}} & \qw   & \qw & \qw &	\cdots & & \qw & \qw & \ctrl{1}&  \qw &  \qw& \qw &	\cdots & &    \qw
    \\
    \lstick{\text{target}} &\qw / & \gate{\text{Layer}\log(L)} & \qw &	\cdots & & \gate{\text{Layer }2} & \gate{\text{Layer }1} & \gate{ih^1(\phi_1)} &  \gate{\text{Layer }1^\dagger} &  \gate{\text{Layer }2^\dagger} & \qw &	\cdots & &    \gate{\text{Layer}\log(L)^\dagger}
      }};

 \node[right] at (-6,-1.5) { \Qcircuit @C=1.4em @R=1.2em {
& \gate{\text{Layer }1}  & \qw & \push{ = \quad  \; \; }  & \gate{h^1h^2(\frac{1}{2}\phi_2)}  & \qw
}};

 \node[right] at (-6,-2.5) { \Qcircuit @C=1.4em @R=1.2em {
& \gate{\text{Layer }2}  & \qw & \push{ = \quad  \; \; }  & \gate{h^1h^3(\frac{1}{2}\phi_3)}  & \gate{h^2h^4(\frac{1}{2}\phi_4)} & \qw }};

 \node[right] at (-6,-3.5) {   \Qcircuit @C=1.4em @R=1.2em {
& \gate{\text{Layer }3}  & \qw & \push{ = \quad  \; \; }  & \gate{h^1h^5(\frac{1}{2}\phi_5)}  & \gate{h^2h^6(\frac{1}{2}\phi_6)} & \gate{h^3h^7(\frac{1}{2}\phi_7)}  & \gate{h^4h^8(\frac{1}{2}\phi_8)} & \qw
}};
\end{tikzpicture}
 \par }
 \caption{Antisymmetric gadget to implement the linear combination of pairwise anticommuting strings  controlled on a single qubit. The gate labeled $ih^1(\phi_1)$ denotes a rotation $\exp (i h^1 \phi_1)$ and the gates $h^jh^k(\phi)$ with the argument $\phi$ signify  rotations $ \mathrm{exp}(h^jh^k\phi)$. \textbf{(a)} Main circuit featuring $\log L$ subroutines referred to as `layers' and their inverses, sandwiching the initial rotation around $h^1$. The control relay is build around the fact that  when the initial rotation is removed, the remaining circuit acts trivially. \textbf{(b)} Layer subroutines featuring more and more rotations $h^j h^k$, that all pairwise commute. }  \label{fig:parallel}
\end{figure}

  While the symmetric gadget has a depth of $O(L)$, the antisymmetric one can achieve a depth of  up to $O(\log \Lambda)$.  We here present it in its  utmost parallel version. The gadget consists of a $h^1$ rotation  sandwiched by $L-1$ rotations on one side and the same amount of counter rotations on the other, see Figure \ref{fig:parallel}. We are going to group those rotations into layers, where rotations inside the same layer commute.  The first layer is a rotation of the string $h^1h^2$ about the angle $\frac{1}{2} \phi_2$. Note that the product $h^1h^2$ is just another Pauli string with imaginary coefficient. Embracing the $h^1$ rotation with the first layer and its inverse yields
  \begin{align}
  &\mathrm{exp}\left(-\frac{1}{2}\phi_2\,h^1h^2 \right) \mathrm{exp}\left(i\phi_1h^1 \right)  \mathrm{exp}\left(\frac{1}{2}\phi_2\,h^1h^2 \right)\notag  \\
  & = \cos \phi_1 + i \sin \phi_1 \cdot  h^1 \; \mathrm{exp}\left(\phi_2\,h^1h^2 \right)  \notag\\
  & = \cos \phi_1 + i \sin \phi_1 \cos \phi_2\cdot  h^1 + i \sin \phi_1 \sin \phi_2 \cdot h^2 \, ,
  \end{align}
    where the second line results from the $(h^1 h^2)$ rotation being annihilated on  the constant part of the inner rotation, but the counter rotation being flipped on the $h^1$-part, creating  a conditional $(h^1h^2)$ rotation around the angle $\phi_2$. The outcome is a superposition of the identity, $ih^1$ and $ih^2$. The second layer will feature rotations of  $h^1 h^3$ and $h^2 h^4$ around new angles $\frac{1}{2}\phi_3$ and $\frac{1}{2}\phi_4$. There, the counter rotation of $h^1 h^3$ will be reversed only  upon encountering $h^1$, and $h^2 h^4$ rotations add up to $\phi_4$ when sandwiching  $h^2$. Note that the $h^1h^3$ and $h^2h^4$ rotations commute and can therefore potentially be executed at the same time.  The result is again a superposition ${i h^k}$ (for all ${k\in [4]})$ and $\mathbb{I}$. The procedure is repeated with every new layer employing rotations of $h^j h^k$ of all different  $j$ and $k$, where all $j$ are drawn from the pool of strings already used in an inner layer, and $h^k$ is a completely new string not yet in the superposition. The third layer would for instance feature rotations of $h^1 h^5$,  $h^2h^6$, $h^3h^7$ and $h^4 h^8$. Since the pool of already used rotations doubles with every layer,  we only need $\log L$  layers in total. A classical procedure  matching the angles $\lbrace \phi_k \rbrace$   with the coefficients $\lbrace \beta_k \rbrace$ is sketched in the Appendix \ref{sec:match}.  \\
The resulting circuit has an easy relay point for the controlled application:  as can be inspected in Figure \ref{fig:parallel}, rotations and counter rotations  cancel each  other when pulling the $h^1$ rotation from the center.\\

   \section{Proof}
      \label{sec:proof}
   In this section, we will prove the results \eqref{eq:results}, for which we will first take a  deeper look into the qubitization mechanism as defined by the unitaries \eqref{eq:old_unitaries1}-\eqref{eq:old_unitaries3} in  \cite{low2019hamiltonian}. \\
    We start with the state $\ket{G}\otimes\ket{\varphi}_\ph$, where $\ket{\varphi}_\ph$ is an ansatz for an energy eigenstate $\ket{E}_\ph$. Let $-1<E<1$ be the corresponding eigenvalue $H \ket{E}_\ph = E \ket{E}_\ph$. Using  $\widehat{V} \cdot \widehat{V} = \mathbb{I}$,  the operator $\widehat{V}$ spans a subspace for every state $\ket{E}_\ph$:
     \begin{align}\label{eq:state1}
   \ee \; &= \; \ket{G}\otimes \ket{E}_\ph \, ,\\ \label{eq:state2}
    \eperp \; &= \; \frac{\widehat{V}-E}{\sqrt{1-E^2}}\ket{G}\otimes \ket{E}_\ph \, .
   \end{align}
   In the basis of $\ee$ and $\eperp$, $\widehat{V}$ has the matrix representation
   \begin{align}
   \left[ \begin{matrix}
   E & \sqrt{1-E^2} \\
   \sqrt{1-E^2} & -E
   \end{matrix} \right]
   \end{align}
    with which alone we cannot extract information about $E$ yet, since it has  eigenvalues $\pm 1$.  However, the operator $\widehat{S}$ acts on the basis of this subspace as  $\left[\begin{smallmatrix} 1 \\    & -1   \end{smallmatrix} \right]$, and so the sequence of $\widehat{S}\widehat{V}$ is represented by the matrix     \begin{align}\label{eq:final_matrix}
   \left[ \begin{matrix}
   E & \sqrt{1-E^2} \\
   -\sqrt{1-E^2} & E
   \end{matrix} \right]  \, .
    \end{align}

The  eigenvalues of this matrix are $E\pm i \sqrt{1-E^2}$, and so we have completely described  the qubitization routine. We now have to show that the same is true when $\widehat{S}$ and $\widehat{G}$ are defined as \eqref{eq:results}.  In particular, we want to do this independently of the representation chosen for the strings $\gamma_j^x$, $\gamma_j^y$ and therefore only  assume   $\gamma^x_k \ket{\bbs{0}} = -i \gamma^y_k \ket{\bbs{0}} = \ket{\bbs{\mu_k}}$.
It immediately  follows that $\widehat{G}$ of \eqref{eq:results} prepares the state $\ket{G}$, and so we only have to show that $\widehat{S}$ of \eqref{eq:results} fulfills $\widehat{S}\ee = \ee$ and $\widehat{S}\eperp = - \eperp$.\\
   We first examine the product $g^y g^x = i\widehat{S}$:
  \begin{align}
  \label{eq:prepare_proof}
    g^y \cdot g^x \;=\;  \sum_{m\in [\Lambda]} \alpha_m {  \gamma^y_m \gamma^x_m} \; +  \;    \frac{1}{2}\sum_{k\in[\Lambda]} \; \sum_{j \in [k-1]}  \sqrt{\alpha_j \alpha_k} \;\left( \gamma^y_j \gamma^x_k + \gamma^y_k \gamma^x_j\right) \, .
    \end{align}
   Furthermore, using the anticommutation relations of the Majoranas we find  $i\gamma^y_k \gamma^x_k \ket{\bbs{\mu_j}} = (-1)^{\delta_{jk}}\ket{\bbs{\mu_j}} $ and
   \begin{align}   \label{eq:prepare_proof2}
    -\frac{i}{2}\left( \gamma^y_j \gamma^x_k + \gamma^y_k \gamma^x_j\right) \ket{\bbs{\mu_m}} \; = \; -\frac{i}{2}\, \gamma^y_j\gamma_k^x\,  \left(1-(-1)^{\delta_{mj}+\delta_{mk}}\right)\ket{\bbs{\mu_m}} \; = \; \begin{cases} \ket{\bbs{\mu_k}} & \text{if}\;\;  m=j \\ \ket{\bbs{\mu_j}} & \text{if} \; \; m=k \\ 0 & \text{else.} \end{cases} \,
\end{align}
Therefore, we can split the product $\widehat{S}\widehat{V}\ee$ into two parts:
    \begin{align}  \label{eq:proof1}
\widehat{S} \widehat{V} \ee  \; \;
& = \; -  i  g^y \cdot g^x \sum_k \sqrt{\alpha_k}\ket{\bbs{\mu_k}}\otimes p^k_\ph \ket{E}_\ph \\
& = \;- \sum_k \overbrace{\left[\sum_m  (-1)^{\delta_{mk}}\,  \alpha_m\right]}^{1-2\alpha_k} \notag \sqrt{\alpha_k} \ket{\bbs{\mu_k}}\otimes p^k_\ph \ket{E}_\ph
\\ &  \quad \; \, + 2 \sum_k  \underbrace{\left[\sum_{j\neq k}  \sqrt{\alpha_j} \label{eq:proof2} \ket{\bbs{\mu_j}}\right]}_{\ket{G}-\sqrt{\alpha_k} \ket{\bbs{\mu_k}}} \otimes \;   \alpha_k \;  p^k_\ph \ket{E}_\ph   \\
&  = \; (2E-\widehat{V}) \ee \, . \label{eq:proof3}
    \end{align}
  Plugging \eqref{eq:proof3} into \eqref{eq:state2} and considering  also that  $\widehat{S} \ee = -i g^y \ket{\bbs{0}}\otimes \ket{E}_\ph  = \ee$, we  verify that $\widehat{S}$ acts  as $Z$ operator onto the subspace spanned by  $\ee$ and $\eperp$.\\

\section{Logarithmic depth}
\label{sec:logdepth}

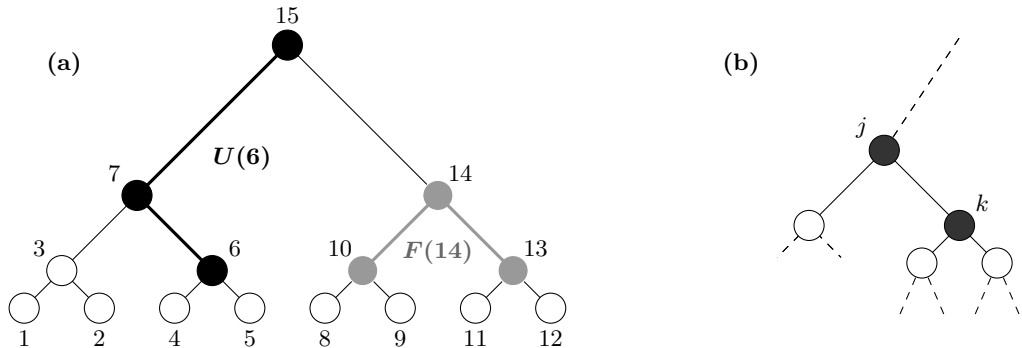
\begin{figure}
{ \centering
\begin{tikzpicture}
\node[] at (-3,1.5){\textbf{(a)}};
\node[] at (6,1.5){\textbf{(b)}};
\node[] at (0,0) {
\begin{tikzpicture}

\draw[] (-1.5,0)--(-1,0.5)--(0,1.5)--(2, 3.5)--(4, 1.5)--(5,.5)--(5.5,0);
\draw[] (0,1.5)--(1,.5)--++(.5,-.5);
\draw[] (1,.5)--++(-.5,-.5);
\draw[] (-.5,0)-- ++(-.5,.5);
\draw[] (2.5,0)--(4,1.5);
\draw[] (5,.5)--++(-.5,-.5);
\draw[] (3.5,0)-- ++(-.5,.5);
\draw[fill=black] (2,3.5) circle[radius=0.2];
\node[] at (2,3.5+.4){15};

\draw[fill=white] (-1.5,0) circle[radius=0.2];
\node[] at (-1.5,-.4){1};
\draw[fill=white] (-.5,0) circle[radius=0.2];
\node[] at (-.5,-.4){2};
\draw[fill=white] (.5,0) circle[radius=0.2];
\node[] at (.5,-.4){4};
\draw[fill=white] (1.5,0) circle[radius=0.2];
\node[] at (1.5,-.4){5};
\draw[fill=white] (-1,.5) circle[radius=0.2];
\node[] at (-1-.3,.5+.3) {3};
\draw[fill=black] (1,.5) circle[radius=0.2];
\node[] at (1+.3,.5+.3){6};
\draw[fill=black] (0,1.5) circle[radius=0.2];
\node[] at (0-.3,1.5+.3){7};
\draw[fill=white] (2.5,0) circle[radius=0.2];
\node[] at (2.5,-.4){8};
\draw[fill=white] (3.5,0) circle[radius=0.2];
\node[] at (3.5,-.4){9};
\draw[fill=white] (4.5,0) circle[radius=0.2];
\node[] at (4.5,-.4){11};
\draw[fill=white] (5.5,0) circle[radius=0.2];
\node[] at (5.5,-.4){12};

\draw[white, fill=black!40] (3,.5) circle[radius=0.2];
\node[] at (3-.3,.5+.3){10};
\draw[white,fill=black!40] (5,.5) circle[radius=0.2];
\node[] at (5+.3,.5+.3){13};
\draw[white,fill=black!40] (4,1.5) circle[radius=0.2];
\node[] at (4+.3,1.5+.3){14};
\draw[very thick] (2,3.5)--(0,1.5)--(1,.5);
\draw[very thick, black!40] (3,.5)--(4,1.5)--(5,.5);
\node[black!60] at (4,.75) {$\boldsymbol{F(14)}$};
\node[black] at (1.4,2) {$\boldsymbol{U(6)}$};
\end{tikzpicture}};
\node[] at (8,0){
\begin{tikzpicture}
\draw[dashed] (0,1.5)-- ++(2*.5,2*.75);
\draw[dashed] (-1.5,0)--(-1,.5)--(-.5,0);
\draw[dashed] (0,1.5)-- ++(2*.5,2*.75);
\draw[dashed] (-1.5,0)--(-1,.5)--(-.5,0);
\draw[dashed] (.5,0) --++(.3 ,-.7);
\draw[dashed] (.5,0) --++(-.3 ,-.7);
\draw[dashed] (1.5,0) --++(.3 ,-.7);
\draw[dashed] (1.5,0) --++(-.3 ,-.7);
\draw[] (-1,.5)--(0,1.5)--(1,.5)--(1.5,0);
\draw[] (.5,0)--(1, .5);
\draw[white, fill=white] (-1.5,0) circle[radius=0.1];

\draw[white, fill=white] (-.5,0) circle[radius=0.1];

\draw[fill=white] (.5,0) circle[radius=0.2];

\draw[fill=white] (1.5,0) circle[radius=0.2];

\draw[fill=white] (-1,.5) circle[radius=0.2];

\draw[fill=black!80] (1,.5) circle[radius=0.2];
\node[] at (1+.3,.5+.3){$k$};
\draw[fill=black!80] (0,1.5) circle[radius=0.2];
\node[] at (0-.3,1.5+.3){$j$};

\end{tikzpicture}} ;
\end{tikzpicture} \par }
\caption{Binary-tree encoding of $\gamma_j^x$ and $\gamma^y_j$. \textbf{(a)} Perfect binary tree with $\Lambda=15$ nodes labeled systematically. We highligted examples of flip  and update sets: $F(14)=\lbrace 10, \, 13, \,14 \rbrace$ and $U(6)=\lbrace 6, \, 7, \,15 \rbrace$. \textbf{(b)} Clipping of a binary tree, showing only qubits (nodes)  the products $\gamma_j^x\gamma_k^x$  and $\gamma_j^y\gamma_k^y$ have support on: the  union of sets $F(j)$ and $F(k)$, with $j$ and $k$ highlighted. } \label{fig:tree}
\end{figure}

In this section, we will change the representation of the Majorana strings, and so be able to show that $g^y$ and $g^x$ (and therefore $\widehat{S}$ and $\widehat{G}$) can be implemented in depth $O(\log \Lambda) $. Let us start by properly generalizing unary encodings. \\  Generalized unary encodings can be defined indirectly by  considering the sets $F(j), U(j) \subseteq [\Lambda]$ for every $j\in [\Lambda]$, that we will refer to as flip and update sets of $j$, respectively. These sets are chosen such that $(\bigotimes_{l\in U(k)} X_l )$ and $(\bigotimes_{m\in F(j)} Z_m)$ only anticommute if $j=k$, and the $\gamma$ strings are defined as
\begin{align}
\label{eq:BK}
\gamma_k^x \; = \; \left( \bigotimes_{l\in U(k)} X_l \right) \prod_{j\in [k-1]} \; \bigotimes_{m\in F(j)} Z_m \quad \text{and} \quad \gamma_k^y \; = \; i\left( \bigotimes_{l\in U(k)} X_l \right) \prod_{j\in[k]} \;  \bigotimes_{m\in F(j)} Z_m \, .
\end{align}
This is an idea originating  from fermionic encodings \cite{seeley2012bravyi,steudtner2018fermion}.  \\
In our case, we want the flip and update sets to be defined on a graph.  Let us consider a tree graph with $\Lambda$ nodes, that are uniquely labeled with the integers $[\Lambda]$ in a certain way. We define:
\begin{itemize}
\item $F(j)$ includes $j$ and all the integers given to the direct children of  node $j$ in the tree.
\item $U(k)$ includes $k$ and all the integers given to the ancestors of node $k$, including the root of the tree.
\end{itemize}
Regarding the labeling of the nodes we will at first assume  $\Lambda$ to be  equal to  $2^N-1$ for some integer $N$, and consider  a perfect binary tree as a graph. See Figure \ref{fig:tree}(a) for an example with $\Lambda=15$, in which two of the sets are highlighted.   The next bigger tree with $\Lambda(N+1)$ nodes can always be  obtained  in taking two trees of size $\Lambda(N)$, adding the number $2^N-1$ to the labels of each node in one of the trees and then connecting the roots of both trees with a new node labeled $2^{N+1}-1$. In this way,  a tree of every size $\Lambda(N)$ can iteratively be reached from the one in Figure \ref{fig:tree}(a). In case $\Lambda$ is not a number $2^N-1$ (for an integer $N$), the next biggest tree can be considered and superfluous nodes deleted. \\
Note that if the tree was a Fenwick tree, we would call the encoding a Bravyi-Kitaev transform \cite{bravyi2002fermionic,seeley2012bravyi,havlivcek2017operator}, but we have made a different choice here to keep the number of children on each node constant. This is an important ingredient of the complexity proof that we will now commence. \\
We will show that with \eqref{eq:BK}, the linear combinations $g^x$ and $g^y$ can  be implemented in  $O(\log \Lambda)$ time using antisymmetric gadgets. In the case of $g^x$, the innermost rotation is about the string $\gamma^x_\Lambda$, which happens to be the root of the tree. From there on, pairs of layers are defined with the following scheme: each layer consist of rotations $\gamma^x_j \gamma^x_k$, with  $j$  being on the level last finished, and $k$ being a direct child of $j$ on an unfinished level. The first layer of every pair features all nodes $j$ of the previous level  and is followed by a layer of rotations $\gamma^x_j \gamma^x_{\ell}$, with the same $j$, but where $\ell$ is the respective other child of each $j$. After these two layers one level is completed and the procedure is repeated until the leaves are included. The implementation of $g^y$ is analogous.   Note that this procedure is slightly different from the one outlined in Section \ref{subsec:anti}, as here a new layer  only uses roughly half of the  pool of strings already used  in inner layers. The reason for this is that only those strings are ensured  not to overlap: Bravyi-Kitaev-like encodings are defined such that most of the $Z$ operators in \eqref{eq:BK} cancel and due to the product of $\gamma^x_j \gamma^x_k$ (respectively $\gamma^y_j \gamma^y_k$ )  the  $X$ strings cancel almost completely since $U(k) = U(j) \cup \lbrace k \rbrace $. The only  qubits on which the product is supported are in the union of the flips sets $F(j)$ and $F(k)$, see Figure \ref{fig:tree}(b) for an illustration. Since  the rotations of the same layer have  no overlap, they can be performed in parallel. Furthermore, as the Pauli weight of those operators is  bounded,  every layer is finished  in constant time and  since the number of layers is proportional to the number of levels, of which there are $O(\log \Lambda)$,  the algorithmic depth exhibits the same scaling. \\
With $\bbs{\mu_k}\neq\bbs{e_k}$,  $\widehat{V}$ has to be adapted, as \eqref{eq:shallowV} is no longer valid. Since the basis of the  binary-tree encoding has the form $\bbs{\mu_k}=\sum_{j\in U(k)}\bbs{e_j}\mod 2$, we find that in a binary-tree version of $\widehat{V}$, all strings $p^k_\ph$ are applied conditioned on the joint parity of the qubits in the flip set $F(k)$. Although a bit longer, the depth of this operation can achieve the same scaling as \eqref{eq:shallowV}.
\section{Conclusion}
\label{sec:conclusion}
We have attempted to make the qubitization more elegant and tractable by eliminating higher-level gates in interpreting its unitaries as fermionic operators: for the preparation subroutine  a delocalized fermion is created in an external register of qubits, and the reflection  is implemented by the parity operator of that fermionic mode. For a logarithmic time complexity, no additional qubits are needed, but only a change in the  representation of those fermions is required. However, to harness this time scaling without using swap or sorting networks, a tree-shaped qubit connectivity would be required. More realistically, one could employ  different encodings for a scaling of at least  $O(\sqrt{\Lambda})$  that is, however, resulting in  local gates on a square lattice. In fact, it might be beneficial to use custom tree graphs which can be embedded into the qubit connectivity.\\ We are aware that all unary encodings are resource intensive for problems defined on lattices however, the qubit requirements could be feasible.   For the simulation of different models, the gadgets for the linear combination of anticommuting strings could be used to relax the qubit requirements by banding  anticommuting Hamiltonian terms together.  In fact, this has been the original  idea behind \cite{izmaylov2019unitary,bonet2019nearly} where the authors set out to reduce the number of measurements in VQE experiments. Note however that we only expect substantial reductions in the number of terms when considering problems with highly delocalized fermions, as in those Hamiltonians we expect large sets of pairwise anticommuting strings. Alternatively, the resource requirements could be relaxed by combining the routine $\widehat{V}$ with circuits that (nearly) diagonalize parts of a Hamiltonian \cite{wiesner1996simulations,babbush2018low,huggins2019efficient}. The  linear combination gadgets could also be used for the preparation of trial states in the simulator. For the benefit of both, qubitization and state preparation, the relationship between the presented gadgets and Givens rotations \cite{wecker2015solving,kivlichan2018quantum} should be studied further.
\acknowledgments
We would like to thank TE O’Brien for some interesting discussions about that topic and CWJ Beenakker for his support. MS was supported by the Netherlands Organization for Scientific Research (NWO/OCW) and an ERC Synergy Grant. SW is supported by an NWO VIDI Grant, the NWO Zwaartekracht QSC and an ERC Starting Grant.
\bibliographystyle{unsrt}
\bibliography{qubit}

\begin{thebibliography}{10}

\bibitem{lloyd1996universal}
Seth Lloyd.
\newblock Universal quantum simulators.
\newblock {\em Science}, 273:1073, 1996.

\bibitem{suzuki1990fractal}
Masuo Suzuki.
\newblock Fractal decomposition of exponential operators with applications to
  many-body theories and {M}onte {C}arlo simulations.
\newblock {\em Physics Letters A}, 146:319, 1990.

\bibitem{suzuki1991general}
Masuo Suzuki.
\newblock General theory of fractal path integrals with applications to
  many-body theories and statistical physics.
\newblock {\em Journal of Mathematical Physics}, 32:400, 1991.

\bibitem{low2019hamiltonian}
Guang~Hao Low and Isaac~L Chuang.
\newblock Hamiltonian simulation by qubitization.
\newblock {\em Quantum}, 3:163, 2019.

\bibitem{poulin2018quantum}
David Poulin, Alexei Kitaev, Damian~S Steiger, Matthew~B Hastings, and Matthias
  Troyer.
\newblock Quantum algorithm for spectral measurement with a lower gate count.
\newblock {\em Physical review letters}, 121:010501, 2018.

\bibitem{izmaylov2019unitary}
Artur~F. Izmaylov, Tzu-Ching Yen, Robert~A. Lang, and Vladyslav Verteletskyi.
\newblock Unitary partitioning approach to the measurement problem in the
  variational quantum eigensolver method.
\newblock {\em Journal of Chemical Theory and Computation}, 16(1):190, 2020.
\newblock PMID: 31747266.

\bibitem{bonet2019nearly}
Xavier Bonet-{M}onroig, Ryan Babbush, and Thomas~E O'{B}rien.
\newblock Nearly optimal measurement scheduling for partial tomography of
  quantum states.
\newblock {\em arXiv:1908.05628}, 2019.

\bibitem{poulin2014trotter}
David Poulin, Matthew~B. Hastings, Dave Wecker, Nathan Wiebe, Andrew~C.
  Doberty, and Matthias Troyer.
\newblock The trotter step size required for accurate quantum simulation of
  quantum chemistry.
\newblock {\em Quantum Information {\&} Computation}, 15:361, 2015.

\bibitem{wecker2014gate}
Dave Wecker, Bela Bauer, Bryan~K Clark, Matthew~B Hastings, and Matthias
  Troyer.
\newblock Gate-count estimates for performing quantum chemistry on small
  quantum computers.
\newblock {\em Physical Review A}, 90:022305, 2014.

\bibitem{childs2019faster}
Andrew~M. Childs, Aaron Ostrander, and Yuan Su.
\newblock Faster quantum simulation by randomization.
\newblock {\em {Quantum}}, 3:182, September 2019.

\bibitem{kivlichan2019phase}
Ian~D Kivlichan, Christopher~E Granade, and Nathan Wiebe.
\newblock Phase estimation with randomized {H}amiltonians.
\newblock {\em arXiv:1907.10070}, 2019.

\bibitem{nielsen2000quantum}
Michael~A. Nielsen and Isaac~L. Chuang.
\newblock {\em Quantum Computation and Quantum Information}.
\newblock Cambridge University Press, 2000.

\bibitem{cleve1998quantum}
Richard Cleve, Artur Ekert, Chiara Macchiavello, and Michele Mosca.
\newblock Quantum algorithms revisited.
\newblock {\em Proceedings of the Royal Society of London. Series A:
  Mathematical, Physical and Engineering Sciences}, 454:339, 1998.

\bibitem{kitaev1995quantum}
A~Yu Kitaev.
\newblock Quantum measurements and the {A}belian stabilizer problem.
\newblock {\em arXiv:quant-ph/9511026}, 1995.

\bibitem{low2017optimal}
Guang~Hao Low and Isaac~L Chuang.
\newblock Optimal {H}amiltonian simulation by quantum signal processing.
\newblock {\em Physical review letters}, 118:010501, 2017.

\bibitem{babbush2018encoding}
Ryan Babbush, Craig Gidney, Dominic~W Berry, Nathan Wiebe, Jarrod McClean,
  Alexandru Paler, Austin Fowler, and Hartmut Neven.
\newblock Encoding electronic spectra in quantum circuits with linear {T}
  complexity.
\newblock {\em Physical Review X}, 8:041015, 2018.

\bibitem{bravyi2002fermionic}
Sergey~B Bravyi and Alexei~Yu Kitaev.
\newblock Fermionic quantum computation.
\newblock {\em Annals of Physics}, 298:210, 2002.

\bibitem{seeley2012bravyi}
Jacob~T Seeley, Martin~J Richard, and Peter~J Love.
\newblock The {B}ravyi-{K}itaev transformation for quantum computation of
  electronic structure.
\newblock {\em The Journal of chemical physics}, 137:224109, 2012.

\bibitem{havlivcek2017operator}
Vojt{\v{e}}ch Havl{\'\i}{\v{c}}ek, Matthias Troyer, and James~D Whitfield.
\newblock Operator locality in the quantum simulation of fermionic models.
\newblock {\em Physical Review A}, 95:032332, 2017.

\bibitem{setia2018superfast}
Kanav Setia, Sergey Bravyi, Antonio Mezzacapo, and James~D Whitfield.
\newblock Superfast encodings for fermionic quantum simulation.
\newblock {\em Physical Review Research}, 1:033033, 2019.

\bibitem{steudtner2019quantum}
Mark Steudtner and Stephanie Wehner.
\newblock Quantum codes for quantum simulation of fermions on a square lattice
  of qubits.
\newblock {\em Physical Review A}, 99:022308, 2019.

\bibitem{jiang2018majorana}
Zhang Jiang, Jarrod McClean, Ryan Babbush, and Hartmut Neven.
\newblock Majorana loop stabilizer codes for error mitigation in fermionic
  quantum simulations.
\newblock {\em Phys. Rev. Applied}, 12:064041, 2019.

\bibitem{sawaya2019resource}
Nicolas~PD Sawaya, Tim Menke, Thi~Ha Kyaw, Sonika Johri, Al{\'a}n Aspuru-Guzik,
  and Gian~Giacomo Guerreschi.
\newblock Resource-efficient digital quantum simulation of $ d $-level systems
  for photonic, vibrational, and spin-$ s $ {H}amiltonians.
\newblock {\em arXiv:1909.12847}, 2019.

\bibitem{chowdhury2018improved}
Anirban~Narayan Chowdhury, Yigit Subasi, and Rolando~D Somma.
\newblock Improved implementation of reflection operators.
\newblock {\em arXiv:1803.02466}, 2018.

\bibitem{motzoi2017linear}
Felix Motzoi, Michael~P Kaicher, and Frank~K Wilhelm.
\newblock Linear and logarithmic time compositions of quantum many-body
  operators.
\newblock {\em Physical review letters}, 119:160503, 2017.

\bibitem{hoffman1972generalization}
David~K Hoffman, Richard~C Raffenetti, and Klaus Ruedenberg.
\newblock Generalization of {E}uler {A}ngles to {N}-{D}imensional {O}rthogonal
  {M}atrices.
\newblock {\em Journal of Mathematical Physics}, 13:528, 1972.

\bibitem{steudtner2018fermion}
Mark Steudtner and Stephanie Wehner.
\newblock Fermion-to-qubit mappings with varying resource requirements for
  quantum simulation.
\newblock {\em New Journal of Physics}, 20:063010, 2018.

\bibitem{wiesner1996simulations}
Stephen Wiesner.
\newblock Simulations of many-body quantum systems by a quantum computer.
\newblock {\em arXiv preprint quant-ph/9603028}, 1996.

\bibitem{babbush2018low}
Ryan Babbush, Nathan Wiebe, Jarrod McClean, James McClain, Hartmut Neven, and
  Garnet Kin-Lic Chan.
\newblock Low-depth quantum simulation of materials.
\newblock {\em Physical Review X}, 8:011044, 2018.

\bibitem{huggins2019efficient}
William~J Huggins, Jarrod McClean, Nicholas Rubin, Zhang Jiang, Nathan Wiebe,
  K~Birgitta Whaley, and Ryan Babbush.
\newblock Efficient and noise resilient measurements for quantum chemistry on
  near-term quantum computers.
\newblock {\em arXiv:1907.13117}, 2019.

\bibitem{wecker2015solving}
Dave Wecker, Matthew~B Hastings, Nathan Wiebe, Bryan~K Clark, Chetan Nayak, and
  Matthias Troyer.
\newblock Solving strongly correlated electron models on a quantum computer.
\newblock {\em Physical Review A}, 92:062318, 2015.

\bibitem{kivlichan2018quantum}
Ian~D Kivlichan, Jarrod McClean, Nathan Wiebe, Craig Gidney, Al{\'a}n
  Aspuru-Guzik, Garnet Kin-Lic Chan, and Ryan Babbush.
\newblock Quantum simulation of electronic structure with linear depth and
  connectivity.
\newblock {\em Physical review letters}, 120:110501, 2018.

\end{thebibliography}

\appendix
\section{Matching angles in the antisymmetric gadget}
\label{sec:match}
\begin{figure}

{\centering
\begin{tikzpicture}[scale=1.2]
\draw[] (-4,0)--++(.5,1.5) node[midway, sloped, above] {$\cos\phi_5$} --++(.5,-1.5) node[midway, sloped, above] {$\bbs{\sin\phi_5}$};

\draw[] (-2,0)--++(.5,1.5) node[midway, sloped, above] {$\cos\phi_7$} --++(.5,-1.5) node[midway, sloped, above] {$\sin\phi_7$};

\draw[] (0,0)--++(.5,1.5) node[midway, sloped, above] {$\bbs{\cos\phi_6}$} --++(.5,-1.5) node[midway, sloped, above] {$\sin\phi_6$};

\draw[] (2,0)--++(.5,1.5) node[midway, sloped, above] {$\cos\phi_8$} --++(.5,-1.5) node[midway, sloped, above] {$\sin\phi_8$};

\draw[] (-3.5,1.5) --++(1,1.5) node[midway, sloped, above] {$\bbs{\cos\phi_3}$} --++(1,-1.5) node[midway, sloped, above] {$\sin\phi_3$};

\draw[] (.5,1.5) --++(1,1.5) node[midway, sloped, above] {$\bbs{\cos\phi_4}$} --++(1,-1.5) node[midway, sloped, above] {$\sin\phi_4$};

\draw[] (-2.5,3)  --++(2,1.5) node[midway, sloped, above] {$\bbs{\cos\phi_2}$} --++(2,-1.5) node[midway, sloped, above] {$\bbs{\sin\phi_2}$};

\draw[] (-6.5,4.5)  --++(3,1.5) node[midway, sloped, above] {$\cos\phi_1$} --++(3,-1.5) node[midway, sloped, above] {$\sin\phi_1$};
\draw[very thick] (-3,0) --++(-.5,1.5)--++(1,1.5)--++(2, 1.5) --++ (2, -1.5)--++(-1,-1.5) --++(-.5, -1.5);

\node[fill=white]  at (-4,0){$ih^1$};
\node[fill=white]  at (-3,0){$\bbs{ih^5}$};
\node[fill=white]  at (-2,0){$ih^3$};
\node[fill=white]  at (-1,0){$ih^7$};
\node[fill=white]  at (0,0){$\bbs{ih^2}$};
\node[fill=white]  at (1,0){$ih^6$};
\node[fill=white]  at (2,0){$ih^4$};
\node[fill=white] at (3,0){$ih^8$};

\node[fill=white]  at (-3.5,1.5){$ih^1$};
\node[fill=white]  at (-1.5,1.5){$ih^3$};
\node[fill=white]  at (.5,1.5){$ih^2$};
\node[fill=white]  at (2.5,1.5){$ih^4$};
\node[fill=white] at (-2.5,3)  {$ih^1$} ;
\node[fill=white] at (1.5,3)  {$ih^2$} ;
\node[fill=white] at (-.5,4.5)  {$ih^1$} ;
\node[fill=white] at (-6.5,4.5)  {$\mathbb{I}$} ;
\node[fill=white] at (-3.5,6)  {$\mathbb{I}$} ;

\end{tikzpicture}	\par }
\caption{Matching the sets of angles $\lbrace \phi_k\rbrace$ to the parameters $\lbrace \beta_k\rbrace$ using a tree describing the circuit in Figure \ref{fig:parallel}, where each forking  stands for a rotation (and counter rotation) where the edges are labeled with the corresponding coefficients. The edges, with which the angle $\phi_2$ is inferred are highlighted, since the ratio of the coefficients on the right and left flank must be equal to $\beta_2/\beta_5$.   }\label{fig:matching}
\end{figure}
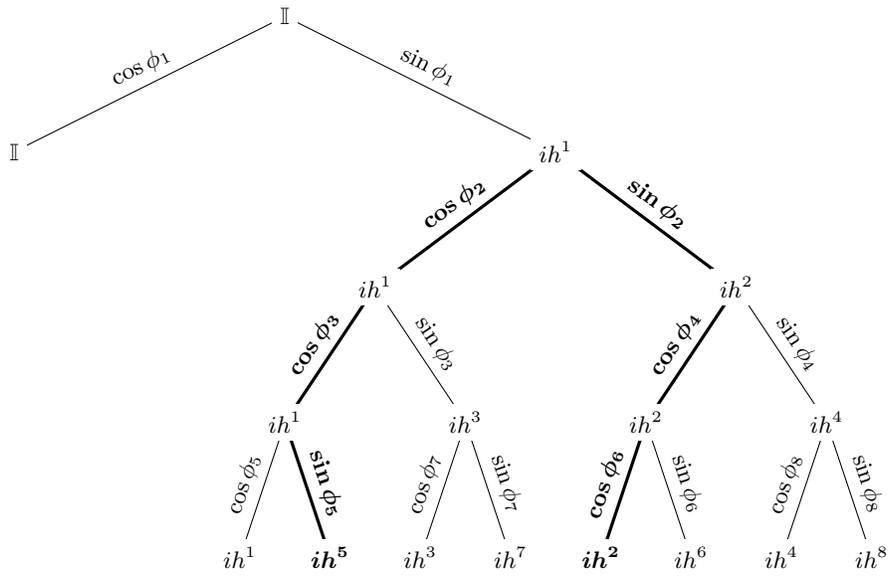

    In this appendix we outline the classical routine  to obtain the set of  angles $\lbrace \phi_k \rbrace$
such that the antisymmetric gadget in Section \ref{subsec:anti} outputs  \eqref{eq:slater_rotation},  given a set of parameters  $\lbrace \beta_k \rbrace$. For this purpose we place all $h^jh^k$ rotations in a tree: the tree of the process described in  Section \ref{subsec:anti} can be found in Figure \ref{fig:matching}. The first level of the tree considers the initial rotation, and, from the inside out, every  rotation and counter rotation of  one of the operators $h^jh^k$ is denoted by a fork of the current node $ih^j$ into  new nodes $i h^j$ and $ih^k$, where the edges are labeled with the corresponding rotation angles.
When the gadget is completed, the tree has nodes labeled with all $\lbrace ih^k\rbrace$  for $k\in [L]$ and with $\mathbb{I}$. The coefficient of one of those operators   is on the one hand  found in \eqref{eq:slater_rotation}  and on the other hand, it can be read out from the tree:  one needs to  find the corresponding operator on the leaf level, and then multiply all the sines and cosines found on edges that connect its ancestors   with each other. For example, in the case of $ih^3$, we need to match the coefficients
\begin{align}
\frac{\sin\theta}{\theta} \beta_3\;  = \; \cos(\phi_7) \,  \sin(\phi_3) \, \cos(\phi_2) \,  \sin(\phi_1)\, .
\end{align}

Considering a pair of operators, the ratio of coefficients in \eqref{eq:slater_rotation} must match the ratio of coefficients obtained from the tree, and so we have a means to infer $\lbrace\phi_k\rbrace$. We start  by comparing the ratios of siblings on the leaf level: $\phi_5$ for instance is inferred by considering the coefficient ratios of $ih^5$ and $ih^1$,
\begin{align}
\frac{\beta_5}{\beta_1} = \tan \phi_5 \, ,
\end{align}
as all other sines and cosines involving the angles $\phi_3$, $\phi_2$ and $\phi_1$ cancel.
 As we have inferred some angles $\lbrace \phi_k \rbrace$ we can consider the coefficient ratios of pairs with the first common ancestor one level lower in the tree and so, level by level,  move towards the root inferring all angles. In Figure \ref{fig:matching}, for instance, we have highlighted the paths used for the ratio of coefficients of  $ih^5$  and $ih^2$, \begin{align}
\frac{\beta_2}{\beta_5} = \frac{\cos(\phi_6) \, \cos (\phi_4)}{\cos(\phi_5) \, \sin (\phi_3)}\tan \phi_2\,  ,
\end{align}
from which $\phi_2$ can be inferred since at that point, the angles $\phi_3$, $\phi_4$, $\phi_5$ and $\phi_6$ are already known.
  The last angle to be matched is $\phi_1$, for which we could make use of the constant coefficient's ratio with  $ih^8$,
\begin{align}
\frac{\theta}{\beta_8} \tan \theta = \tan(\phi_1)\, \sin(\phi_2)\, \sin(\phi_4) \, \sin(\phi_8) \, .
\end{align}
\end{document}